\documentclass[%
 reprint,
 amsmath,amssymb,
 aps,
prb,
]{revtex4-2}
\usepackage[normalem]{ulem}
\usepackage{graphicx}
\usepackage{dcolumn}
\usepackage{physics}
\usepackage{bm}
\usepackage{hyperref}
\usepackage{xcolor}
\usepackage{cancel}

\begin{document}

\preprint{APS/123-QED}

\title{Krylov complexity, mode-resolved complexity and entanglement entropy across phase transitions in the non-Hermitian extended Su-Schrieffer-Heeger model}


\author{Ling-Feng Zhang}
\affiliation{Department of Physics, City University of Hong Kong, Kowloon, Hong Kong, China}
\author{Wing Chi Yu}\email{wingcyu@cityu.edu.hk}
\affiliation{Department of Physics, City University of Hong Kong, Kowloon, Hong Kong, China}

\date{\today}

\begin{abstract}
We investigate phase transitions in the extended Su–Schrieffer–Heeger (SSH) model with next-nearest-neighbor hoppings and an imaginary staggered chemical potential. In the presence of small non-Hermiticity, exceptional points emerge in pairs from the gap-closing momenta near the topological phase boundaries of the Hermitian limit. Utilizing the Krylov spread complexity and entanglement entropy, we analyze two dynamical protocols: (i) preparing the non-Hermitian ground state via a unitary transformation, and (ii) evolving the system under the non-Hermitian Hamiltonian. We show that the spread complexity, and long-time spread complexity as well as entanglement entropy can effectively signal phase transitions in the first and second protocols, respectively. To unravel the detailed structure of the transitions, we introduce the momentum-resolved complexity that identifies the characteristic modes and tracks their evolution with the driving parameter. In the regime where the system possesses a purely imaginary spectrum, we further identify dynamical phases based on the saturation behavior of the spread complexity. The entanglement entropy is also found to exhibit similar saturation behavior, thereby providing a more experimentally accessible probe of the dynamical phases.
\end{abstract}

\maketitle


\section{\label{sec:level1}Introduction}
Understanding quantum phase transitions is a cornerstone of condensed matter physics. Tools from information theory, such as entanglement entropy and quantum information measures, have been successfully employed to characterize these phases and their transitions. With the rapid development of quantum technologies, quantum complexity has emerged as a powerful tool to investigate the dynamics of states or operators in Hamiltonian systems. Intuitively, quantum complexity quantifies the difficulty of preparing a quantum state or simulating a system, reflecting the computational cost of specific tasks on a quantum computer. This concept has found applications across diverse fields, ranging from black holes and quantum gravity in high-energy physics to quantum computing and strongly correlated many-body systems.

Since the landmark work of Ref. \cite{PhysRevX.9.041017}, Krylov complexity has emerged as a key framework for characterizing the information-theoretic aspects of quantum dynamics. Originally introduced to quantify how rapidly an initial operator spreads within the operator space under Heisenberg evolution, it has since been extended to the Schr\"{o}dinger evolution of quantum states, giving rise to the notion of spread complexity \cite{PhysRevD.106.046007}. Compared to standard complexity measures \cite{Quantum_circuit_complexity1993,Haferkamp_2022,dowling2006geometryquantumcomputation,nielsen2005geometricapproachquantumcircuit,doi:10.1126/science.1121541}, Krylov complexity offers a more natural and robust metric \cite{PhysRevResearch.6.L042001,PhysRevLett.132.160402}. For a comprehensive review, see Ref. \cite{Nandy_2025}.

Initial studies on Krylov complexity utilized the saturation plateau as a valid indicator of quantum chaos \cite{Rabinovici:2021qqt,Rabinovici:2022beu}, and recent studies have further explored the diagnostic potential of its pre-saturation behavior \cite{Erdmenger_2023,Camargo_2024,PhysRevB.111.L060202,PhysRevResearch.7.023028,Alishahiha_2025}. The utility of Krylov complexity has since extended far beyond chaos, finding applications in characterizing equilibrium phases, such as topological transition \cite{PhysRevB.106.195125,Caputa:2022yju}, parity and time reversal ($\mathcal{PT}$) transitions in non-Hermitian systems \cite{Bhattacharjee:2022lzy,PhysRevB.111.174207,6lvg-7qdn,PhysRevB.110.064320}, general gap closings \cite{chaudhuri2026krylovcomplexityfidelitysusceptibility}, and quantum many-body scars \cite{PhysRevB.111.165106}. Furthermore, in non-equilibrium settings, it has been utilized to probe weak-ergodicity breaking \cite{caputa2025complexitypxpscarsrevisited}, many-body localization transition \cite{Trigueros:2021rwj,PhysRevB.109.014312,mf2z-mrpd,PhysRevB.110.L180101,ganguli2024spreadcomplexitynonhermitianmanybody}, and non-unitary quantum dynamics \cite{PhysRevResearch.5.033085,PhysRevD.109.L121902,Bhattacharya_2024,Chakrabarti:2025hsb,PhysRevB.111.064203}. It has also been instrumental in identifying novel phases \cite{PhysRevB.111.174207,6lvg-7qdn,xia2025quantumcomplexityphasetransition} and advancing quantum reservoir computing \cite{m9kq-wrln}.
In particular, an executable experimental proposal has recently been put forward in Ref. \cite{_indrak_2024}. 

In recent years, non-unitary dynamics involving dissipation due to coupling to external environments have garnered significant attention \cite{10.21468/SciPostPhysLectNotes.99}. An important class of such dynamics is governed by non-Hermitian Hamiltonians \cite{Ashida02072020}, which arise naturally in open quantum systems, specifically in the no-click limit of monitored evolution \cite{Daley04032014} or in systems with gain and loss \cite{El-Ganainy2018,doi:10.1126/science.aar7709}. The exotic spectral and topological properties \cite{PhysRevX.8.031079,10.21468/SciPostPhys.7.5.069,PhysRevLett.121.086803,RevModPhys.93.015005}, as well as the distinct correlation  and entanglement patterns of these systems \cite{PhysRevLett.120.185301,PhysRevResearch.2.033069,PhysRevLett.130.010401,Zhang2025-ub}, have stimulated extensive investigation. In particular, entanglement dynamics have been shown to exhibit intriguing transitions driven by the non-Hermitian skin effect \cite{PhysRevX.13.021007,PhysRevB.108.214308} or spectral transitions \cite{PhysRevB.108.214308,PhysRevLett.126.170503,PhysRevB.104.L161107,PhysRevB.107.L020403,10.21468/SciPostPhys.14.5.138,PhysRevLett.130.230401,PhysRevResearch.6.013131}. Especially, for non-Hermitian SSH model with gain and loss, Ref. \cite{10.21468/SciPostPhys.14.5.138} shows that the entanglement transitions from volume-volume and volume-area law coincide with the regime where $\mathcal{PT}$-symmetry breaking and fully $\mathcal{PT}$-symmetry breaking occur, respectively. Refs. \cite{PhysRevB.111.174207,6lvg-7qdn} show that spread complexity can detect these $\mathcal{PT}$ transitions and reveal dynamical phases with different saturation patterns in the purely imaginary spectrum regime. The saturation dynamics in these phases are primarily controlled by the slowest decay modes of the spectrum. However, the detailed structure of the transitions and whether other physical quantities that can characterize the dynamical phases in the purely imaginary spectrum regime deserve further investigation.

 While the SSH model has been widely studied, the non-Hermitian physics of its extended version incorporating further-neighbor hoppings remains largely unexplored. In the Hermitian limit, in contrast to the original SSH model, whose phase diagram consists of only a trivial phase and a topological phase with winding number 1, the extended version exhibits a much richer phase diagram hosting multiple topological phases characterized by different winding numbers \cite{Hui2026-yu,Hui2026-in,maffei2018,PhysRevA.95.061601}. Besides, the model also demonstrates intricate non-equilibrium dynamics. A notable example is the emergence of anomalous dynamical quantum phase transitions\cite{PhysRevB.110.054312,Carapeto-Carmelo2025-ye}, a phenomenon absent in the conventional SSH framework. Furthermore, the experimental feasibility of such extended systems has been demonstrated across various platforms, including recent implementations in photonic \cite{Leefmans2022,PhysRevResearch.6.043087}, acoustic \cite{PhysRevApplied.19.054028,PhysRevLett.131.157201} and ultracold atomic systems \cite{Xie2019}.

In this paper, we investigate the phase transitions in the extended version of the SSH model with the next-nearest-neighbor hopping and an imaginary staggered potential. When small non-Hermiticity is turned on, we find that exceptional points emerge from the gap-closing points in the momentum space in the vicinity of phase transition boundaries of the Hermitian model. The number of emergent exceptional points is twice the number of the original gap-closing points, which in turn is equal to the absolute difference in the winding number of the phases on the two sides of the phase boundaries.

Furthermore, we study the dynamical behaviors of the non-Hermitian extended SSH model via the Krylov complexity and the entanglement entropy. Two dynamical protocols which admit analytical solutions are considered \cite{PhysRevB.111.174207,6lvg-7qdn}: (1) unitary evolution of an initial state, which is the lowest-weighted state of $\mathrm{su(2)}$ algebra, into the non-Hermitian Hamiltonian's ground state; and (2) long-time evolution under the non-Hermitian Hamiltonian from the initial state. In both cases, we find that the spread complexity is sensitive to the topological phase transitions in the Hermitian model, and the transitions associated with changes in the number of exceptional points in the non-Hermitian case. However, minor peaks or valleys are observed within a phase in the long-time spread complexity and entanglement entropy under the second protocol, which suggests Krylov complexity in the first protocol may be a better indicator of transitions.

Moreover, motivated by the observation that
the spread complexity depends on contributions from different modes in the momentum space, we introduce the mode-resolved complexity and the corresponding fidelities to reveal the detailed structure of the transitions. The mode-resolved complexity displays clear fingerprints of the transitions, identifies the critical modes responsible for the signatures in the complexity, and tracks their evolutions with the driving parameters. To quantify the similarity of mode-resolved complexity's distribution, we examine several fidelity measures. The fidelity of the mode-resolved complexity across the driving parameter shows abrupt changes across transitions and accurately captures the phase boundaries in both the Hermitian and non-Hermitian models. On the other hand, the fidelity across momentum detects the gap-closing momenta and their evolution with the model parameter in the Hermitian limit, while in the non-Hermitian case, it identifies the characteristic momenta where the onset or offset of gap-closing occurs. 

In addition, we find that the extended SSH model also hosts distinct dynamical phases in the purely imaginary spectral regime, as characterized by the saturation times of the spread complexity and the entanglement entropy. For spread complexity, we derive an analytical saturation time scaling that is dictated by the slowest decay mode. Unlike in the conventional SSH model, the critical mode in the extended model varies continuously with the driving parameter. We further find that the entanglement entropy exhibits a similar saturation relation.

The rest of this paper is organized as follows. In Sec. \ref{sec2}, we introduce the extended SSH model and its equilibrium phase diagrams. In Sec. \ref{sec3}, we study the spread complexity in unitary and non-unitary dynamics. In Sec. \ref{sec4}, we investigate the detailed structure of the phase transitions by mode-resolved complexity. In Sec. \ref{sec5}, the saturation behavior in the regime possessing purely imaginary spectrum is investigated from the perspectives of spread complexity and entanglement entropy. Finally, we summarize the findings in Sec. \ref{sec6}.

\section{The model\label{sec2}}
The Hamiltonian of the extended SSH model with an imaginary staggered chemical potential reads \cite{Maffei_2018,PhysRevB.110.054312,10.21468/SciPostPhys.14.5.138,PhysRevB.97.045106},
\begin{figure}[!tb]
    \centering\includegraphics[width=\columnwidth]{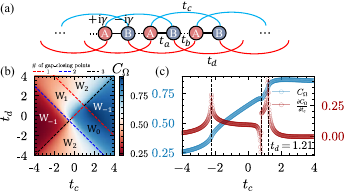}
    \caption{(a) Schematic diagram of the extended SSH chain with the unit cell constructed by site A and site B. The intra-cell (inter-cell) hopping $t_a$ ($t_b$) is denoted by the black solid (dashed) line. The blue and red curves denote the next-nearest-neighbor hopping terms $t_c$ and $t_d$, respectively. The red (blue) circles denote sites with opposite on-site imaginary potential $\pm i\gamma$. (b) The spread complexity of the Hermitian extended SSH model on the $t_c$-$t_d$ plane with $t_a=t_b=-1$. $W_n$ denotes the phase characterized by a winding number of $n$. The red, blue and black dashed lines are the topological phase transition boundaries and the colors indicate the number of gap-closing points. (c) The spread complexity and its derivative over $t_c$ along the line $t_d=1.21$.}
    \label{fig1}
\end{figure}

\begin{align}\label{eq:real_ham}
\begin{split}
    H = &\sum_{j=1}^{L}\Big(t_a c^{\dagger}_{j,A} c_{j,B} + t_b c^{\dagger}_{j,B} c_{j+1,A}\\
    &+ t_c c^{\dagger}_{j,A} c_{j+1,B} + t_d c^{\dagger}_{j,B} c_{j+2,A} + \text{H.c.}\Big)\\
    &+i\gamma \sum_{j=1}^{L}\Big(c^{\dagger}_{j,A} c_{j,A} - c^{\dagger}_{j,B} c_{j,B}\Big),
\end{split}
\end{align}
where $L$ is the number of unit cells, $A$ and $B$ denote the two sublattices, $t_a=-J-h/2$ and $t_b=-J+h/2$ are the intra- and inter-cell hoppings in the original SSH model. The next-nearest-neighbor hoppings that preserve chiral-symmetry $t_c$ and $t_d$ are considered (cf. Fig. \ref{fig1}(a)). In the Hermitian limit, the model has topological phases with winding number $W=1,-1$ and $2$ respectively, and a trivial phase with winding number $W=0$ (cf. Fig. \ref{fig1}(b)). The model has $\mathcal{PT}$-symmetry, where $\mathcal{P}$ is the parity operator defined as $\mathcal{P}c_{j,A}\mathcal{P}=c_{L-j+1,B}$, $\mathcal{P}c_{j,B}\mathcal{P}=c_{L-j+1,A}$ and $\mathcal{T}$ is the anti-unitary time-reversal operator defined as $\mathcal{T}i\mathcal{T}=-i$. Hereafter, we set $J=1$ as the energy unit.

The model can be solved exactly via the Fourier transformation as \cite{6lvg-7qdn} 
\begin{equation}
    \left(\begin{array}{c}
        c_{j,A}\\
        c_{j,B}  
    \end{array}\right)=\frac{e^{-i\frac{\pi}{4}\sigma_x}}{\sqrt{L}}\sum_{k\in\mathcal{K}}e^{ikj}\left(\begin{array}{c}
        a_{k}\\
        b_{k}  
    \end{array}\right),
\end{equation}
where $\sigma_x$ is a Pauli matrix, and we use anti-periodic boundary conditions, hence the $\mathcal{K}$ satisfy
\begin{equation}
    \mathcal{K}=\{k_n=\frac{2\pi}{L}(-\Big\lceil \frac{L-1}{2}\Big\rceil +n+\frac{1}{2})\mid n=0,1,...,L-1\},
    \label{eq:kn}
\end{equation}
which brings the Hamiltonian of the system into the form of 
\begin{equation}\label{eq:k_ham}
    H = \sum_{k}\Psi_{k}^{\dagger}H(k)\Psi_{k},
\end{equation}
where $\Psi_k = (a_k,b_k)^{T}$ and $H(k)$ is the Bloch Hamiltonian $H(k) = \vec{d}(k)\cdot \vec{\sigma}$ where the associated Bloch vector has the components
\begin{align}
    \begin{split}
        d_x(k) &= t_a+(t_b+t_c)\cos{k}+t_d\cos{(2k)},\\
        d_y(k) &= -i\gamma,\\
        d_z(k) &=(t_b-t_c)\sin{k}+t_d\sin{(2k)},
    \end{split}
\end{align}
and $\vec{\sigma}$ is the Pauli matrix vector.

The Hamiltonian can be diagonalized by  $V_k^{-1}\vec{d}(k)\cdot\vec{\sigma}V_k=\text{diag}[\Lambda(k),-\Lambda(k)]$, where
\begin{equation}
    V_k = \left(\begin{array}{cc}
        u_k & u_k \\
         v_k^+& v_k^- 
    \end{array}\right)\equiv \left(\begin{array}{cc}
        \vec{v}_{+}(k) & \vec{v}_{-}(k)
    \end{array}\right),
\end{equation}
and 
\begin{equation}
    \vec{v}_{\pm}(k) = \frac{1}{\sqrt{2d(k)(d(k)\mp d_z(k))}}\left(\begin{array}{c}
        d_x(k)-i d_y(k)\\
        \pm d(k)-d_z(k)
    \end{array}\right),
\end{equation}
with $d(k)=\sqrt{d_x^2(k) + d_y^2(k) + d_z^2(k)}$. The Bogoliubov quasiparticles can be obtained by
\begin{equation}
  \left(\begin{array}{cc}
        \bar{\eta}_{+,k} & \bar{\eta}_{-,k}
    \end{array}\right) := \Psi_k^\dagger V_k = \left(\begin{array}{cc}
        u_k a_k^\dagger+v_k^{+} b_k^\dagger & u_k a_k^\dagger + v_k^{-}b_k^\dagger
    \end{array}\right),
\end{equation}
and 
\begin{equation}
  \left(\begin{array}{c}
        \eta_{+,k} \\
        \eta_{-,k}
    \end{array}\right) := V_k^{-1}\Psi_k = \frac{1}{\det{V_k}}\left(\begin{array}{c}
        v_k^{-} a_k-u_k b_k\\
        -v_k^{+} a_k + u_k b_k
    \end{array}\right),
\end{equation}
where $\eta_{\pm,k}$ are non-Hermitian quasiparticle fermionic operators obeying the anticommutation relations
\begin{equation}
    \{\bar{\eta}_{s,k},\eta_{s',k'}\}=\delta_{ss'}\delta_{kk'},\quad \{\eta_{s,k},\eta_{s',q'}\} = \{\bar{\eta}_{s,k},\bar{\eta}_{s',k'}\}=0,
\end{equation}
and $\eta_{s,k}^\dagger\neq\bar{\eta}_{s,k}$. Hence, $H$ can be written in diagonal form as
\begin{equation}
    H = \sum_{k\in\mathcal{K}} \Lambda(k)(\bar{\eta}_{+,k}\eta_{+,k}-\bar{\eta}_{-,k}\eta_{-,k}),
\end{equation}
where
\begin{equation}\label{eq:spectrum}
    \Lambda(k) = \pm d(k)\equiv \pm E(k)\pm i\Gamma(k),
\end{equation}
where $E(k):= \Re(\Lambda(k))$ and $\Gamma(k):= \Im(\Lambda(k))$. For a given $k\in\mathcal{K}$, $\Lambda(k)$ is either real or imaginary.

\begin{figure}[!tb]
    \centering\includegraphics[width=\columnwidth]{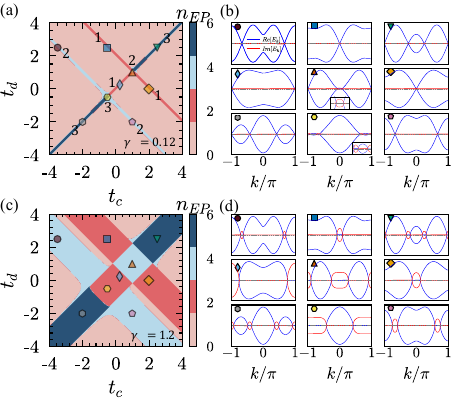}
    \caption{Distribution of exceptional points in the $t_c$-$t_d$ plane for (a) weak ($\gamma = 0.12$) and (c) moderate ($\gamma = 1.2$) non-Hermiticity. The markers lie at the phase transition boundaries of the Hermitian extended SSH model ($\gamma = 0$). The numbers adjacent to these markers denote the number of gap-closing points in momentum space. Panels (b) and (d) display the real (blue curves) and imaginary (red curves) parts of the energy spectra in momentum space, corresponding to the parameters indicated by markers in panels (a) and (c), respectively. The gray dotted line represents the zero-energy reference level ($E=0$). The insets in the bottom two panels of the middle column in (b) show a zoom-in of the vicinity around the exceptional points.}
    \label{fig2}
\end{figure} 

In the Hermitian limit $(\gamma=0)$, at the phase transition boundaries, the number of gap-closing points is equal to the absolute difference of the winding numbers of the adjacent phases (cf. Fig. \ref{fig1}(b) and Fig. \ref{fig2}(a)). Upon introducing weak non-Hermiticity ($\gamma=0.12$), exceptional points emerge from these gap-closing points (cf. Fig. \ref{fig2}(b)), with the number of exceptional points being twice that of the original gap-closing points. As the non-Hermiticity increases to a moderate strength ($\gamma=1.2$), the exceptional points regime expands (cf. Fig. \ref{fig2}(c)); however, some exceptional points eventually coalesce and vanish (cf. Fig. \ref{fig2}(d)). In the strong non-Hermitian limit, this correspondence no longer holds as the entire spectrum becomes purely imaginary.

We further show that $H(k)$ is an element of semisimple Lie algebra $\mathrm{su(2)}\times \mathrm{su(2)}$ and derive its ground state following Ref. \cite{6lvg-7qdn}. Equation (\ref{eq:k_ham}) can be rewritten as
\begin{equation}
H=\sum_{\substack{k\in\mathcal{K}^{+}\\s=\pm}} H^s(k),
\end{equation}
where $\mathcal{K}^{+}=\{k\in\mathcal{K}\mid k>0\}$ and
\begin{equation}
    H^s(k) = 2d_z(k)J_z^s(k) + d_{+}^s(k)J_{+}^s(k)+d_{-}^s(k)J_{-}(k),
\end{equation}
where 
\begin{align}
    \begin{split}\label{eq:su_algebra}
        J_{+}^{+}(k) &:= a_k^\dagger b_k,\\
        J_{-}^{+}(k) &:= b_k^\dagger a_k,\\
        J_{z}^{+}(k) &:= \frac{1}{2}(a_k^\dagger a_k - b_k^\dagger b_k),\\
        J_{+}^{-}(k) &:= b_{-k}^\dagger a_{-k},\\
        J_{-}^{-}(k) &:= a_{-k}^\dagger b_{-k},\\
        J_{z}^{-}(k) &:= -\frac{1}{2}(a_{-k}^\dagger a_{-k} - b_{-k}^\dagger b_{-k}),
    \end{split}
\end{align}
and note that $d_{+}^{+}(k)=d_{-}^{-}(k)=d_x(k)-i d_y(k)$ and $d_{-}^{+}(k)=d_{+}^{-}(k)=d_x(k)+i d_y(k)$. In this form, for each momentum $k\in{\mathcal{K}_+}$, $H^{+}(k)+H^{-}(k)$ can be seen as an element of the semisimple Lie algebra $\mathrm{g}=\mathrm{g}^{+}\times\mathrm{g}^{-}=\mathrm{su(2)}\times\mathrm{su(2)}$. We have the following commutators:
\begin{equation}
    \begin{split}[J_{+}^{s}(k),J_{-}^{s'}(k')]&=2\delta_{ss'}\delta_{kk'}J_z^s(k),\\
    [J_{z}^{s}(k),J_{\pm}^{s'}(k')]&=\pm\delta_{ss'}\delta_{kk'}J_{\pm}^s(k).
    \end{split}
\end{equation}
The superscript $s$ indexes the Lie algebra, and the subscripts $\pm, z$ denote the corresponding generators.

We consider a reference state
\begin{equation}\label{eq:gs}
    \ket{\psi_{\text{ref}}} = \prod_{k\in\mathcal{K}^{+}}a_{-k}^\dagger b_k^\dagger\ket{0}=\otimes_{k\in\mathcal{K}}\ket{1/2,-1/2}_k.
\end{equation}
For each $k$, $\ket{1/2,-1/2}_k$ is a lowest-weight state of $\mathrm{su(2)}$, satisfying
\begin{align}
    \begin{split}
        J_{-}^\pm(k)\ket{1/2,-1/2}_{\pm k}&=0,\\
        J_{+}^\pm(k)\ket{1/2,-1/2}_{\pm k}&=\ket{1/2,1/2}_{\pm k},\\
        J_{z}^\pm(k)\ket{1/2,\pm1/2}_{\pm k}&=\pm\frac{1}{2}\ket{1/2,\pm1/2}_{\pm k}.
    \end{split}
\end{align}

The un-normalized ground state of the Hamiltonian is
\begin{widetext}
\begin{align}
\begin{split}
    \ket{\widetilde{\text{GS}}}&=\prod_{k\in\mathcal{K}^+}\bar{\eta}_{-,-k} \bar{\eta}_{-,k}\ket{0}\\
    &=\prod_{k\in\mathcal{K}^+}u_{-k}v_{k}^{-}\left(a_{-k}^\dagger b_{k}^\dagger +\frac{u_k}{v_k^{-}}a_{-k}^\dagger a_{k}^\dagger + \frac{v_{-k}^{-}}{u_{-k}}b_{-k}^\dagger b_{k}^\dagger + \frac{u_k v_{-k}^{-}}{u_{-k}v_{k}^{-}}b_{-k}^\dagger a_k^\dagger \right)\ket{0}.
\end{split}
\end{align}
\end{widetext}
Using Eq. (\ref{eq:su_algebra}) and then normalizing, it reads
\begin{align}
    \begin{split}
        \ket{\text{GS}}&=\Big\|\ket{\widetilde{\text{GS}}}\Big\|^{-1}\ket{\widetilde{\text{GS}}}\\
        &=\prod_{k\in\mathcal{K}^+}\frac{u_{-k}v_{k}^{-}}{|u_{-k}v_{k}^{-}|}\frac{\exp{\tau^{+}_k J_{+}^{+}(k)+\tau^{-}_k J_{+}^{-}(k)}}{\sqrt{(1+|\tau^{+}_k|^2)(1+|\tau^{-}_k|^2)}}\ket{\psi_{\text{ref}}},
    \end{split}
\end{align}
with $\tau^{+}_k=u_k/v_k^{-}$ and $\tau^{-}_k=v_{-k}^{-}/u_{-k}$.

\section{spread complexity in the unitary and non-unitary dynamics}\label{sec3}
In this section, we generalize the derivation in Ref. \cite{6lvg-7qdn} to the extended SSH model. We investigate the spread complexity for a state that evolves unitarily to the ground state of the Hamiltonian $H$, and non-unitarily under the Hamiltonian, respectively. Spread complexity becomes analytically solvable when the Hamiltonian is an element of a Lie algebra and the initial state $\ket{\psi_{\text{ref}}}$  is the lowest-weight state of the algebra. 

\subsection{Unitary evolution from the reference state to the Hamiltonian's ground state}

To obtain the ground state $\ket{\text{GS}}$ from the reference state via unitary evolution at time $t=1$, we construct a Hamiltonian
\begin{equation}
    H_\Omega = \sum_{k\in\mathcal{K}^+,s=\pm}{H}^s_\Omega(k),
\end{equation}
where $H^s_\Omega(k)=(i\alpha^{s}_k J^{s}_{+}(k)+\text{H.c.})$ with $\alpha^{s}_k=\exp{i\arg{(\tau^s_k)}}\arccos{(1+|\tau^s_k|^2)^{-1/2}}$. Utilizing the normal-ordering $\mathrm{su(2)}$ decomposition formula \cite{Ban:93}, the unitary operator for each mode reads
\begin{widetext}
    \begin{equation}
    \begin{split}
        \exp{-iH^s_\Omega(k)}&=\exp{\alpha^s_k J_{+}^s(k) -(\alpha^s_k)^* J_{-}^s(k)}\\
        &=\exp{\tau_k^s J_{+}^s(k)}\exp{\ln(1+|\tau_k^s|^2)J_z^s(k)}\exp{-(\tau_k^s)^* J_{-}^s(k)}.
    \end{split}
\end{equation}
\end{widetext}

\begin{figure}[!tb]
    \centering\includegraphics[width=\columnwidth]{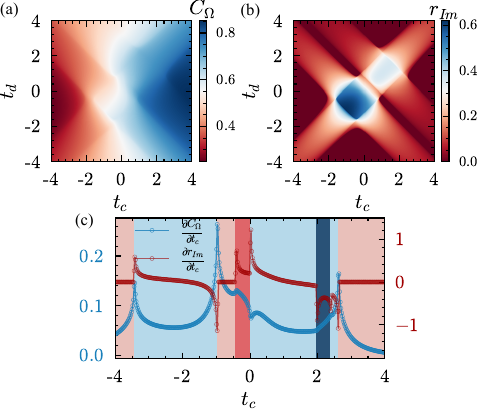}
    \caption{(a) The spread complexity of the non-Hermitian extended SSH model with an imaginary staggered chemical potential $\gamma = 1.2$ on the $t_c$-$t_d$ plane. (b) The ratio of imaginary eigenvalues over the whole spectrum. (c) The derivative of spread complexity (the ratio of imaginary eigenvalues) over $t_c$ is denoted by the blue (red) line with circles, along the line $t_d=1.21$. The background denotes the number of exceptional points [see Fig. \ref{fig2}(c)].}
    \label{fig3}
\end{figure}

Hence, the unitary evolution state reads
\begin{equation}
    \ket{\Omega}=\prod_{\substack{k\in\mathcal{K}^+\\s=\pm}}\ket{\Omega_k^s}=\prod_{\substack{k\in\mathcal{K}^+\\s=\pm}}\frac{\exp{\tau_{k}^s J_+^s(k)}}{\sqrt{1+|\tau_k^s|^2}}\ket{\psi_{\text{ref}}}.
\end{equation}
Up to a phase $\prod_{k\in\mathcal{K}^+}\frac{u_{-k}v_k^{-}}{|u_{-k}v_k^{-}|}$, $\ket{\Omega}$ is equivalent to $\ket{\text{GS}}$, and is a generalized coherent state of $\ket{\psi_{\text{ref}}}$ defined in the coset space $\mathrm{SU(2)}^{\otimes L}/\mathrm{U(1)}^{\otimes L}$ \cite{Perelomov_1986,Kam_Zhang_Feng_2023}.  
The spread complexity of the ground state is given by:
\begin{align}
\begin{split}
    C_\Omega&:=\lim_{L\rightarrow\infty}\frac{1}{L}\sum_{\substack{k\in\mathcal{K^{+}}\\s=\pm}} C^s_{\Omega}(k;t=1),
\end{split}
\end{align}
where
\begin{eqnarray}
C^s_{\Omega}(k;t=1)= \braket{1/2,1/2}{\Omega^s_k} = \frac{|\tau_k^s|^2}{1+|\tau_k^s|^2}
\label{eq:Comegak}
\end{eqnarray}
is the complexity of the corresponding $k$ mode. Note that the phase is irrelevant in the calculation of the spread complexity.

In the Hermitian limit, while $C_\Omega$ is continuous across the $t_c$-$t_d$ plane (cf. \ref{fig1}(b)), its derivative with respect to $t_c$ (or $t_d$) displays discontinuities or divergences at the topological phase transition points (cf. Fig. \ref{fig1}(c) for an example). The abrupt changes in the spread complexity and its derivative can be attributed to the discontinuous changes in $C^s_{\Omega}(k,t=1)$ at the gap-closing momenta. For a purely real spectrum,
\begin{eqnarray}
C^{\pm}_{\Omega}(k;t=1)=\frac{[d_x(k)\mp\gamma]^2}{[E(k)+d_z(k)]^2+[d_x(k)\mp\gamma]^2},
\end{eqnarray}
which diverges at $E(k)=0$ and $\gamma=0$.

As the imaginary staggered potential is turned on, the spread complexity becomes more complex (cf. Fig. \ref{fig3}(a)). The imaginary spectrum affects the dynamics of the system. To investigate the influence of spectral properties on spread complexity, we consider the ratio of imaginary eigenvalues (cf. Fig. \ref{fig3}(b)) in the whole spectrum $r_{Im}$ and the number of exceptional points $n_{EP}$ (cf. Fig. \ref{fig2}(c)). We find that the boundaries at which the two quantities change most sharply are consistent with each other. The derivative of the spread complexity and the ratio of imaginary spectrum with respect to $t_c$ (or $t_d$)  exhibit non-analytic behavior at where $n_{EP}$ changes its value (cf. Fig. \ref{fig3}(c)). The observed sensitivity of spread complexity to gap-closing points in both Hermitian and non-Hermitian regimes is consistent with the results reported in Ref. \cite{chaudhuri2026krylovcomplexityfidelitysusceptibility}.

\subsection{Non-unitary evolution of the reference state}
For non-unitary evolution, with the normal-ordering $\mathrm{su(2)}$ decomposition, we can obtain the evolution operator for each mode:
\begin{widetext}
    \begin{align}
    \begin{split}
         \exp{-iH^s(k) t}&=\exp{-it\left(2d_z(k)J_z^s(k) + d_{+}^s(k)J_{+}^s(k)+d_{-}^s(k)J_{-}^s(k)\right)} \\
         &=\exp{A_{+}^{s}(k;t)J_{+}^s(k)}\exp{\ln{(A_z^s(k;t))J_z^s(k)}}\exp{A_{-}^s(k;t)J_{-}^s(k)},
    \end{split}
\end{align}
\end{widetext}
where 
\begin{equation}
    A_{\pm}^s(k;t)=-i\frac{d_{\pm}^s(k)}{\Lambda(k)}\frac{\sin{(\Lambda(k)t})}{\cos{(\Lambda(k)t)}+i\frac{d_{z}(k)}{\Lambda(k)}\sin{(\Lambda(k)t)}}
\end{equation}
and
\begin{equation}
    A^s_z = [\cos{(\Lambda(k)t)}+i\frac{d_z(k)}{\Lambda(k)}\sin{(\Lambda(k)t)}]^{-2}.
\end{equation}
\begin{figure}[!tb]
    \centering\includegraphics[width=\columnwidth]{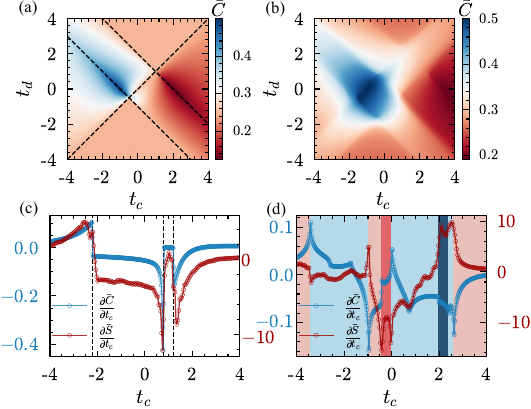}
    \caption{(a),(b) The long-time spread complexity on $t_c$-$t_d$ plane for the Hermitian and non-Hermitian extended SSH model with $\gamma=1.2$, respectively. (c),(d) The derivative of long-time complexity and half-chain entanglement entropy with respect to $t_c$ for the Hermitian and non-Hermitian extended SSH model along the line $t_d=1.21$, respectively. The background denotes the number of exceptional points. }
    \label{fig4}
\end{figure} 
The evolution state is a generalized coherent state of the reference state. Hence, up to a phase factor, the normalized non-unitary evolved state reads
\begin{equation}
    \ket{\psi(t)}=\prod_{\substack{k\in\mathcal{K}^+\\s=\pm}}\frac{\exp{A_{+}^s(k;t)J^s_+(k)}}{\sqrt{1+|A_{+}^s(k;t)|^2}}\ket{\psi_{\text{ref}}}.
\end{equation}
In this case, the mode-resolved complexity reads
\begin{eqnarray}
C^s(k;t)= \frac{|A^s_{+}(k;t)|^2}{1+|A^s_{+}(k;t)|^2}.
\label{Eq:Cbark}
\end{eqnarray}
To investigate the system's dynamics, we estimate the asymptotic long-time spread complexity $\bar{C}$ by averaging over a suitable time window:
\begin{equation}\label{eq:time_average}
    \bar{C} = \frac{1}{t_f-t_0}\int_{t_0}^{tf} C(t) dt,
\end{equation}
where $[t_0,t_f]$ is an appropriate time window in which the behavior of $C(t) = \lim_{L\rightarrow\infty}\frac{1}{L}\sum_{\substack{k\in\mathcal{K^{+}}\\s=\pm}} C^s(k;t)$ has reached stationary. 

The long-time spread complexity $\bar{C}$ exhibits similar features as $C_{\Omega}$ in the Hermitian case (cf. Fig. \ref{fig4}(a) and Fig. \ref{fig1}(b)) and in the non-Hermitian case  (cf. Fig. \ref{fig4}(b) and Fig. \ref{fig3}(a)). The derivative of $C_{\Omega}$ and $\bar{C}$ both show abrupt changes at the phase transitions (cf. Figs. \ref{fig1}(c) and \ref{fig4}(c)), and Figs. \ref{fig3}(c) and \ref{fig4}(d)). However, there are non-trivial signatures in the first derivative of $\bar{C}$ observed within a phase that do not correspond to a transition in terms of $n_{EP}$. This suggests $C_{\Omega}$ may be a better indicator of transition than $\bar{C}$, unlike in the conventional SSH model in which they work equally well \cite{6lvg-7qdn}.

\begin{figure*}[!tb]
    \centering\includegraphics[width=.88\textwidth]{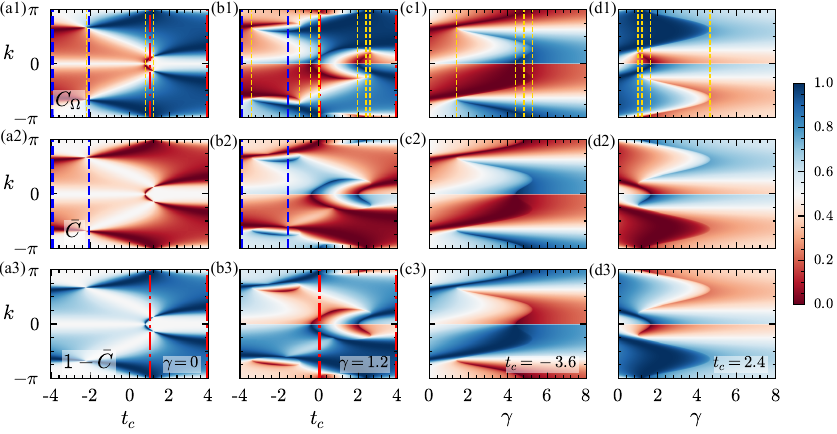}
    \caption{The top, middle, and bottom panels display the momentum-space distributions of $C_\Omega$, $\bar{C}$ and $1-\bar{C}$ for $t_d=1.21$, respectively. The yellow dashed lines in (a1) indicate the gap-closing points, and in (b1, c1, d1) indicate the parameters in which $n_{EP}$ changes. In the first column, the blue (red) dashed lines denote the boundaries where $C_\Omega\approx\bar{C}$ ($C_\Omega\approx 1-\bar{C}$). In the second (third) column, $C_\Omega \approx \bar{C}$ ($C_\Omega\approx 1-\bar{C}$).} 
    \label{fig5}
\end{figure*}

\section{The mode-resolved complexity}\label{sec4}
In this section, we investigate the mode-resolved complexity in Eq. (\ref{eq:Comegak}) and Eq. (\ref{Eq:Cbark}) to uncover the detailed structures that are typically averaged out by a simple summation over $k$ in the spread complexity. This investigation is particularly relevant to translationally invariant free-fermion models, wherein a global quench decouples the non-equilibrium dynamics into mutually independent momentum sectors.

We begin by the distribution of the mode-resolved complexity. Figure \ref{fig5}(a1) shows the distribution of the mode-resolved complexity in Eq. (\ref{eq:Comegak}) as a function of $t_c$ at $t_d=1.21$ in the Hermitian limit. In the figure, $C_\Omega^+$ and $C_\Omega^-$ are shown for $k>0$ and for $k<0$, respectively. The plot clearly displays a fingerprint of the topological phase transitions, originating from the gap-closing critical momenta. As non-Hermiticity is turned on, for $\gamma=1.2$, the mode-resolved complexity in Fig. \ref{fig5}(b1) displays finerprints that correspond to phases with different $n_{EP}$ (ref. Fig. \ref{fig3}(c)). In contrast to the Hermitian case where the mode contribution to the complexity is symmetric about $k=0$, the gapless modes with opposite momenta contribute anti-symmetrically about 1/2 due to $C^{\pm}_\Omega(k)=1/2\mp d_x(k)/(2\gamma)$. For example, focusing on the regime $-3\lessapprox t_c\lessapprox -1$, the system has four gapless modes. The complexity corresponds to the gapless mode with the most positive $k$, let's denote it as $k_{EP}$, is close to zero, suggesting the corresponding mode in the target ground state is nearly locked to that in the initial reference state. In contrast, the gapless mode at the corresponding negative $|k_{EP}|$ has a complexity close to one, suggesting its mostly spread. Figures \ref{fig5}(c1) and (d1) show the mode-resolved complexity at different values of $t_c$ when $\gamma$ varies. Besides exhibiting signatures about the phase transition, the mode-resolved complexity further decodes the information on how the gapless modes evolve when the non-Hermiticity strength increases.

The middle panel in Figure \ref{fig5} shows the long-time average of the mode-resolved complexity in Eq. (\ref{Eq:Cbark}). Similar to the momentum-space distributions of $C_\Omega$ (cf. the top panel in Fig. \ref{fig5}), the long-time average of the mode-resolved complexity also exhibit signatures of phase transitions. These signatures are evident at gap-closing points or the exceptional points, where abrupt changes in the mode-resolved $C_\Omega$ is observed. Furthermore, the distribution of the mode-resolved $\bar{C}$ resembles that of $C_\Omega$. $C_\Omega$ is close to $1-\bar{C}$ or $\bar{C}$ due to the similar momentum-space distributions in both Hermitian (cf. Fig.~\ref{fig5}(a1)-(a3)) or non-Hermitian regimes (cf. Fig.~\ref{fig5}(b1)-(b3)).

\begin{figure*}[!tb]
    \centering\includegraphics[width=.88\textwidth]{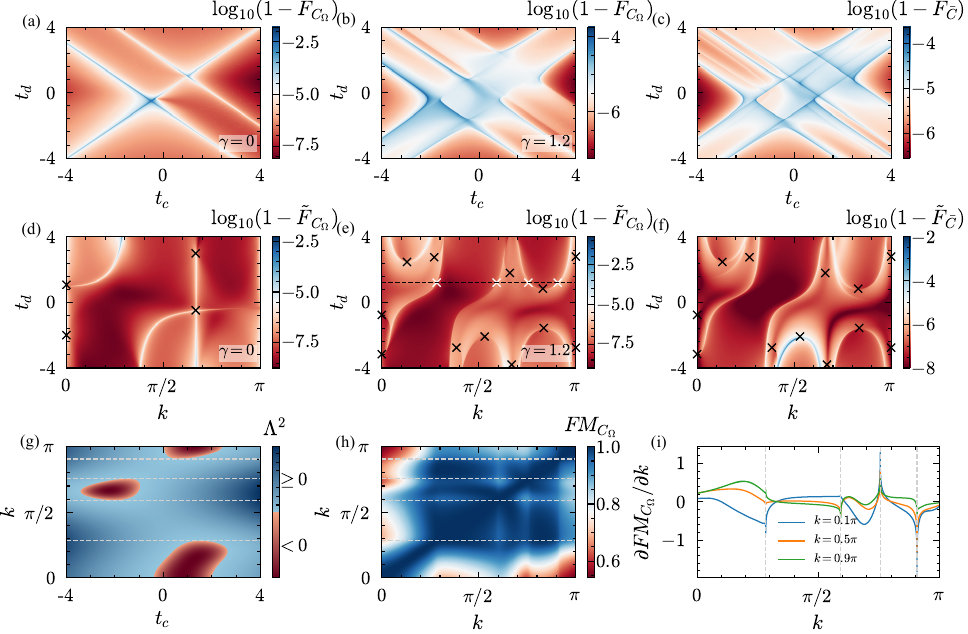}
    \caption{The top and middle panels display the fidelity of the mode-resolved complexity as a function of the parameter $t_c$ and momentum $k$, respectively. (a)(d) depict the fidelity based on the mode-resolved complexity $C_\Omega(k)$ for the Hermitian case, while (b)(e) and (c)(f) depict the fidelities based on $C_\Omega(k)$ and $\bar{C}(k)$ for the non-Hermitian case. White cross markers in (e)  indicate examples of the gap closing-points for $t_c\in[-4,4]$ at where $\tilde{F}(k)$ changes sharply. The black dashed line at $t_d=1.21$ denotes the parameter range investigated in (g)-(i).  Black cross markers highlight the specific values of $t_d$ and $k$ at which a sharp change of $\tilde{F}_{C_{\Omega}}$ or $\tilde{F}_{\bar{C}}$ appears or disappears, signaling the emergence or disappearance of phases encountered within $t_c \in [-4, 4]$, during the sweep of the parameter $t_d$. (g) The energy spectrum in Eq. (\ref{eq:spectrum}) as a function of $k$ and $t_c$ at $t_d=1.21$. (h) Fidelity map over $k$ based on $C_\Omega$. (i) The gradient profiles for three representative $k$ values in the fidelity map in (g). The gray dashed lines across (g)-(i) are aligned with each other and with the $k$ values of the white cross markers in (e). Other parameters are $t_a=t_b=-1$.}
    \label{fig6}
\end{figure*} 

To quantity the similarity in the distribution of the mode-resolved complexity, we introduce the fidelity based on it. The fidelity can be defined over $t_c$ with fixed momenta $k\in[0,\pi]$ as
\begin{eqnarray}
F_{C_\Omega}(t_c)=\left|\frac{\braket{v(t_c)}{v(t_c+\delta)}}{\braket{v(t_c)}{v(t_c)}\braket{v(t_c+\delta)}{v(t_c+\delta)}}\right|,
\end{eqnarray}
where $\ket{v(t_c)}=[C^+_\Omega(t_c,k=0),...,C^+_\Omega(t_c,k=\pi)]^T$ and $\delta$ is a small increment in $t_c$. In an analogous fashion, we can also define the mode-resolved fidelity $\tilde{F}_{C_{\Omega}}(k)$, which is evaluated across $k\in[0,\pi]$ with fixed parameter $t_c$ in an interval $[-4,4]$, i.e.
\begin{eqnarray}
\tilde{F}_{C_\Omega}(k)=\left|\frac{\braket{\tilde{v}(k_n)}{\tilde{v}(k_{n+1})}}{\braket{\tilde{v}(k_n)}{\tilde{v}(k_n)}\braket{\tilde{v}(k_{n+1})}{\tilde{v}(k_{n+1})}}\right|,
\end{eqnarray}
where $\ket{\tilde{v}(k_n)}=[C^+_\Omega(t_c=-4,k_n),...,C^+_\Omega(t_c=4,k_n)]^T$ and $k_n$ is the discretized positive momentum in Eq. (\ref{eq:kn}). Similarly, we can also define the fidelities $F_{\bar{C}}(t_c)$ and $\tilde{F}_{\bar{C}}(k)$ based on the mode-resolved long-time complexity $\bar{C}$. The top panel in Figure \ref{fig6} shows the fidelity $F_{C_{\Omega}}$ and $F_{\bar{C}}$ over $t_c$ for the Hermitian and non-Hermitian cases for $t_d\in[-4,4]$. The phase boundaries are completely captured by the sharp changes in both fidelities (cf. Figs. \ref{fig1}(d) and \ref{fig2}(c)). 

The middle panel of Fig. \ref{fig6} displays the mode-resolved fidelities $\tilde{F}_{C_{\Omega}}(k)$ and $\tilde{F}_{\bar{C}}(k)$. In the Hermitian case as shown in Fig. \ref{fig6}(d), it depicts the movement of gap-closing momenta along $t_d$ and implies phase transitions. The gap-closing points within $t_c\in[-4,4]$ are manifested as sharp changes in the mode-resolved fidelity. In the non-Hermitian case, the sharp changes in the mode-resolved fidelity in Figs. \ref{fig6}(e) and (f) instead capture the momenta extrema of the gap-closing regions, and how they evolve along $t_d$. As a demonstrative example, let us consider the case of $t_d= 1.21$, the characteristic momenta extrema $k^c$'s are indicated by white cross markers in Fig. \ref{fig6}(e). The energy spectrum $\Lambda^2$ in Eq. (\ref{eq:spectrum}) at the corresponding parameters is shown in Fig. \ref{fig6}(g), with the $k^c$'s located by the gray dashed lines. The red regions correspond to the modes that have purely imaginary energies whereas the blue regions correspond to real energies. For $0<k<k^c_1\approx 0.3\pi$, gap-closing modes emerge at some range of $t_c$ located at the boundaries of the red regions. Since the gapless modes depend on the driving parameter $t_c$, they give rise to a relatively large variation in the mode-resolved complexity, therefore resulting in the relatively low fidelity in Fig. \ref{fig6}(e). On the other hand, for $k^c_1<k<k^c_2\approx 0.6\pi$, the spectrum is gapped for all $t_c\in[-4,4]$. The mode-resolved complexity varies smoothly and therefore has a relatively high mode-resolved fidelity in the corresponding regime in Fig. \ref{fig6}(e). Similar arguments follow for other $k^c$'s.

Furthermore, we can also quantify the similarities among the mode-resolved complexities within the parameter range $t_c \in [-4, 4]$ using a fidelity map\cite{PhysRevB.104.075142}
\begin{eqnarray}
FM_{C_{\Omega}}(k,k')=\left|\frac{\braket{\tilde{v}(k_n)}{\tilde{v}(k_{m})}}{\braket{\tilde{v}(k_n)}{\tilde{v}(k_n)}\braket{\tilde{v}(k_{m})}{\tilde{v}(k_{m})}}\right|,
\end{eqnarray}
which takes into account the overlap between two $|\tilde{v}\rangle$'s at $k_n, k_m\in[0,\pi]$ instead of just the neighboring $k_n$'s as in $\tilde{F}_{C_{\Omega}}$. Again, taking an example in the non-Hermitian case (cf. the black dashed line in Fig. \ref{fig6}(e)), Fig. \ref{fig6}(h) shows the corresponding fidelity map and Fig. \ref{fig6}(i) shows its derivative along some selected one-dimensional k-path. High fidelity is predominantly observed among modes sharing resembling features, thereby partitioning the momentum space into multiple distinct sectors. These transitions, as indicated by the gray dashed lines, align well with the characteristic momenta $k_c$'s where the onset/ offset of gap closing occurs.

In addition, the migration of the gap-closing points or the characteristic momenta extrema $k^c$'s signals the emergence or disappearance of phases. We take the example in the Hermitian limit (cf. Fig \ref{fig6}(d)), starting from the initial parameter value of $t_d = -4$, the system undergoes two phase transitions for $t_c\in[-4,4]$ (c.f. Fig. \ref{fig1}(b)) and the corresponding gap-closing momenta are reflected by the abrupt changes in $\tilde{F}_{C_{\Omega}}(k)$ in Fig. \ref{fig6}(d). An increase in $t_d$ toward $-2$ causes the trajectories of these gap-closing momenta to migrate continuously within the momentum space, and the overall features of the fidelity remain qualitatively unchanged. However, at the critical threshold of $t_d = -2$, a new gap-closing point emerges in the vicinity of $k = 0$. This feature signals a transition between the $W_0$ and $W_{-1}$ regimes (cf. Fig. \ref{fig1}(b)). As shown in Fig. \ref{fig6}(e) and (f), when turning on the non-Hermiticity, the emergency/disappearance of most of the phase transitions (cf. Fig. \ref{fig2}(c)) as we scan $t_d$ are also captured by both fidelities based on $C_\Omega$ and $\bar{C}$. However, certain transitions in the low-$|t_d|$ regime remain difficult to resolve due to the vanishingly small fidelity in these regions.

\section{saturation behavior in the regime with purely imaginary spectrum}\label{sec5}
In this section, we study the saturation time $t^*$ of the spread complexity and entanglement entropy in the non-unitary dynamics of the non-Hermitian extended SSH model.

\subsection{Saturation of spread complexity}\label{subsec:Complexity}
In the regime where the Hamiltonian possesses a purely imaginary spectrum, i.e., $\Lambda(k)=\pm i\Gamma(k)$, to study the saturation behavior, it can be defined a spread fidelity \cite{PhysRevB.111.174207,6lvg-7qdn} : 
\begin{equation}\label{eq:spread_fid}
    |\Delta C(t)| = |C(t)-C(t\rightarrow\infty)|,
\end{equation}
where $C(t\rightarrow\infty)=1/2$ as $L\rightarrow\infty$ in this regime. We define a numerical saturation time $t(\epsilon)$ as the time when the inequality $|\Delta C(t)| < \epsilon$ is satisfied.

The spectrum of $H$ (cf. Eq. (\ref{eq:spectrum})) can be expanded as
\begin{equation}
    \Lambda^2(k) = A\cos^3{k}+B\cos^2{k}+C\cos{k}+D-\gamma^2,
\end{equation}
where
\begin{align}
\begin{split}
    A &= 8 t_c t_d,\\
    B &= 4 (t_a t_d + t_b t_c),\\
    C &= 2 (t_b t_d +  t_a t_b + t_a t_c - 3 t_c t_d),\\
    D &= t_a^2 + t_d^2 + (t_b-t_c)^2 - 2 t_a t_d.
\end{split}
\end{align}
The relaxation behavior is determined by the slowest decay mode. For a purely imaginary spectrum $\Lambda(k)=i\Gamma(k)$, $\Gamma^2(k) = \gamma^2 -(A\cos^3{k}+B\cos^2{k}+C\cos{k}+D)$. With $\frac{d\Gamma^2(k)}{dk}=0$ and $\frac{d^2\Gamma^2}{dk^2}>0$, the minimum $\Gamma^2(k^{*})$ can be found. The Taylor expansion of $\Gamma(k)$ around the $k^*$  to the leading order in $k^2$ is  
\begin{equation}
    \Gamma(k) \approx \Gamma(k^*)+\frac{1}{2}T^{''}(k-k^*)^2,
\end{equation}
where
\begin{widetext}
    \begin{equation}
    \Gamma(k^*) = \sqrt{\gamma^2-[(t_a+(t_b+t_c)\cos{k^*}+t_d\cos{(2k^*)})^2+((t_b-t_c)\sin k^*+t_d \sin{(2k^*)})^2]},
\end{equation}
and $\Gamma^{''}(k^*)=\frac{d^2 \Gamma(k)}{dk^2}\Big|_{k^*}$.
\end{widetext}
The spread fidelity (cf. Eq. (\ref{eq:spread_fid})) can be expanded as \cite{PhysRevB.111.174207,6lvg-7qdn} 
\begin{align}
\begin{split}
    |\Delta C(t)| &= \left|\sum_{s=\pm}\int_{0}^{\pi}\frac{|A_{+}^{s}(k;t)|^2}{1+|A_{+}^{s}(k;t)|^2}\frac{dk}{2\pi}-\frac{1}{2}\right|\\
    &\approx \left|\int_{0}^{\pi}\frac{-A_1(k;t) A_2(k)-A^2_1(k;t)}{\gamma^2 A_2(k)\left(1+\frac{2A_1(k;t)}{A_2(k)}+\frac{A_1^2(k;t)}{\gamma^2 A_2(k)}\right)}\frac{dk}{2\pi}\right|\\
    &\approx \int_{0}^{\pi} \frac{A_1(k;t)}{\gamma^2}\frac{dk}{2\pi},
\end{split}
\end{align}
where $A_1(k;t) = 2\Gamma^2(k)\exp{-2\Gamma(k)t}$ and $A_2(k)=\gamma^2-d_x^2(k)$. In the limit $t\rightarrow\infty$, the approximation in the second line is obtained by taking $\coth^2(\Gamma(k)t) \approx 1+4e^{-2\Gamma(k)t}$. Subsequently, the final expression is evaluated to leading order in $A_1$.
The integral can be further simplified by employing a saddle-point approximation around $k^*$:
\begin{align}
\begin{split}
    |\Delta C(t)|&\approx \int_{0}^{\pi}\frac{2\Gamma^2(k)}{\gamma^2}\exp{-2\Gamma t}\frac{dk}{2\pi}\\
    &\approx \frac{\Gamma^2(k^{*})}{\pi\gamma^2} e^{-2\Gamma(k^*)t}\int_{-\infty}^{+\infty}  dk \exp{-\Gamma^{''}(k^*)(k-k^*)^2 t}\\
    &\approx \mathcal{C}_1\frac{e^{-2\Gamma(k^*)t}}{\sqrt{t}},
\end{split}
\end{align}
where $\mathcal{C}_1=\frac{\Gamma^2(k^{*})}{\gamma^2\sqrt{\pi\Gamma^{''}(k^*)}}$ denotes the prefactor. By imposing the condition $|\Delta C(t)|<\epsilon$, we can determine the time $t(\epsilon)$ at which the inequality is satisfied. Specifically, 
\begin{equation}
    |\Delta C(t)| \approx \mathcal{C}_1\frac{e^{-2\Gamma(k^*)t}}{\sqrt{t}}< \epsilon.
    \label{Eq:epsilonC1}
\end{equation}
By squaring both sides and rearranging terms, Equation (\ref{Eq:epsilonC1}) can be rewritten as:
\begin{equation}
    t e^{4\Gamma(k^*)t}>\frac{\mathcal{C}_1^2}{\epsilon^2}.
\end{equation}
Let $z=4\Gamma(k^*)t$,
\begin{equation}
    ze^{z} = \frac{4\Gamma(k^*)\mathcal{C}_1^2}{\epsilon^2}.
\end{equation}
Thus, \begin{equation}
    z = W_0(\frac{4\Gamma(k^*)\mathcal{C}_1^2}{\epsilon^2}),
\end{equation}
where $W_0(x)$ is the Lambert W function. We have
\begin{equation}
    t(\epsilon) \simeq \frac{1}{4\Gamma(k^*)}W_0(\frac{4\Gamma(k^*)\mathcal{C}_1^2}{\epsilon^2}).
\end{equation}
Since $W_0(x\rightarrow\infty) \simeq \ln{x} - \ln{(\ln{x})}$, for small $\epsilon$,
\begin{equation}\label{eq:t_epsilon}
    t_\epsilon \simeq \frac{1}{4\Gamma(k^*)}\Big[\ln(\frac{4\Gamma(k^*)\mathcal{C}_1^2}{\epsilon^2})-\ln(\ln(\frac{4\Gamma(k^*)\mathcal{C}_1^2}{\epsilon^2}))\Big]. 
\end{equation}
which still relates to the prefactor $C_1$. For $\epsilon\rightarrow0$,  the $\epsilon$-independent saturation time reads:
\begin{equation}\label{eq:t_epsilon_indep}
    t^* = \lim_{\epsilon\rightarrow0}\frac{t_\epsilon}{\ln{1/\epsilon}}=\frac{1}{2\Gamma(k^*)}.
\end{equation}
The saturation time is related to the relaxation timescale, i.e., the inverse of the gap. This $\epsilon$-independent description defines the stationary state obtained by times $t\gg t^{*}$
and also provides a tool for characterizing the relaxation dynamics.

\begin{figure}[!tb]
    \centering\includegraphics[width=\columnwidth]{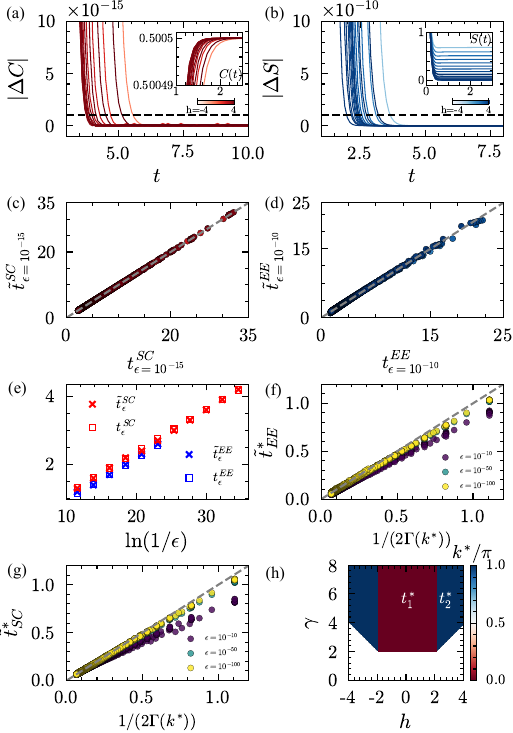}
    \caption{(a),(b) The dynamics of the spread complexity and entanglement entropy and their distance to saturation value, along the line $\gamma=4.86$, respectively. The color denotes the value of $h$, with the deeper shade indicating the larger value. The dashed black lines are $|\Delta C|=10^{-15}$ and $|\Delta S|=10^{-10}$, respectively. (c),(d) The linear relation between the numerical saturation time $\tilde{t}_{\epsilon}$ obtained from spread complexity (SC) and entanglement entropy (EE), respectively, and the analytical saturation time $t_{\epsilon}$ defined by Eq. (\ref{eq:t_epsilon}) for $\epsilon=10^{-15}$ and $\epsilon=10^{-10}$, respectively. The color denotes the value of $h$. (e) $\tilde{t}_{\epsilon}$ agrees well with the time $t_{\epsilon}$ obtained from Eq. (\ref{eq:t_epsilon}) for all $\epsilon$ considered. (f),(g) The linear relation between the $\epsilon$-independent saturation time $t^{\ast}_{SC/EE}=\tilde{t}_{\epsilon}^{SC/EE}/\ln(1/\epsilon)$ and the analytical $t^{\ast}$ defined by the slowest decay mode in Eq. (\ref{eq:t_epsilon_indep}). The color denotes the value of $\epsilon$.  (h) The slowest decay mode $k^{\ast}$ on the $h$-$\gamma$ plane of non-Hermitian SSH model ($t_c=t_d=0$). The white regions correspond to the points where the spectrum is either purely or partially real.}
    \label{fig7}
\end{figure} 

\begin{figure}[!tb]
    \centering\includegraphics[width=\columnwidth]{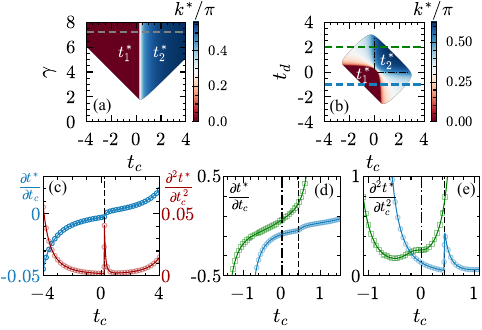}
    \caption{(a) The slowest decay mode $k^{\ast}$ in the $t_c-\gamma$ plane for non-Hermitian SSH model with $h=0$ and $t_d=0$ and (b) in the $t_c-t_d$ plane with $h=0$ and $\gamma=4$. The black dash-dotted lines denote the $t_2^*$ solution corresponding to the case where $k^{\ast}=\arccos{(-\frac{t_c+t_d-1}{4(t_c +t_d)})}$. The gray dashed line in (a) and the blue and green dashed lines in (b) represent the parameter ranges investigated in (c) and in (d) and (e), respectively. The $t_1^{\ast}$ and $t_2^{\ast}$ denote the two different saturation behaviors. (c)-(e) The first and second derivatives of saturation time $t^{*}$ with respect to $t_c$. The black dashed and dash-dotted lines indicate the transition points and the crossover regime, respectively.}
    \label{fig8}
\end{figure}

Let us begin with the non-Hermitian SSH model $(t_c=t_d=0)$.  We numerically pinpoint the saturation time of the spread complexity with the condition $\epsilon=10^{-15}$. The saturation time $\tilde{t}_{\epsilon=10^{-15}}^{SC}$ exhibits a non-monotonic behavior as hopping parameter $h$ increases from $-4$ to $4$, first decreasing from a large value to a minimum, and then increasing again. Furthermore, the values of saturation time are more concentrated for intermediate values of $h$ compared to those at sufficiently large/small values of $h$, suggesting a distinct dependence on $h$ (cf. Fig. \ref{fig7}(a)). We verify that $\tilde{t}_{\epsilon=10^{-15}}^{SC}$ is almost close to the analytical result $t_{\epsilon=10^{-15}}$ obtained from Eq. \ref{eq:t_epsilon} (cf. Fig. \ref{fig7}(c)).  Moreover, these results are consistent with the variation of $\epsilon$ (cf. Fig. \ref{fig7}(e)). By extrapolating the  saturation time $\tilde{t}_{\epsilon}^{SC}$ to the $\epsilon\rightarrow 0$ limit, we find that the numerically obtained $\epsilon$-independent saturation time $\tilde{t}^{\ast}_{SC}\equiv\tilde{t}_{\epsilon}^{SC}/\ln{(1/\epsilon)}$ approaches the analytical result $t^{\ast}$ obtained from Eq. (\ref{eq:t_epsilon_indep}) (cf. Fig. \ref{fig7}(g)). Thus, the slowest decay mode $k^{\ast}$ plays an important role in saturation behavior. The $k^{\ast}$ can only be $0$ or $\pi$, suggesting $t^*$ has two behaviors:
\begin{equation}
    \begin{cases}
    t_1^{\ast}=\frac{1}{2\sqrt{\gamma^2-4}}, & \text{if $k^{\ast}=0$}.\\
    t_2^{\ast}=\frac{1}{2\sqrt{\gamma^2-h^2}}, & \text{ if $k^{\ast}=\pi$}.
  \end{cases}
\end{equation}
The first derivative of $t^{\ast}$ with respect to
$h$ is discontinuous, indicating the presence of the
dynamical phase transition, which is consistent with that reported in Ref. \cite{6lvg-7qdn} (cf. Fig. \ref{fig7}(h)).

We then consider the parameters with $h=0$ and $t_d=0$, and investigate in the purely imaginary spectrum in the $t_c-\gamma$ plane (cf. Fig. \ref{fig8}(a)). The $t^{*}$ now becomes
\begin{equation}
  \begin{cases}
    t_1^{\ast}=\frac{1}{2\sqrt{\gamma^2-t_c^2+4 t_c-4}}, & \text{if $k^{\ast}=0$}.\\
    t_2^{\ast}=\frac{1}{2\sqrt{\gamma^2-t_c^2-(1-t_c)^2/(4 t_c)-2 t_c -2}}, & \text{if $k^{\ast}=\arccos{(\frac{1-t_c}{4t_c})}$}.
  \end{cases}
\end{equation}
The first and second derivatives of $t^{\ast}$ with respect to $t_c$ exhibit a transition signal when the saturation time changes from $t_1^{\ast}$ to $t_2^{\ast}$ (cf. Fig. \ref{fig8}(c)). Unlike in the original non-Hermitian SSH model, in which $k^*$ is independent of the driving parameter in each of the dynamical phases and the saturation time plateaus in one phase, $t^*$ in the extended SSH model varies continuously with the driving parameter in both phases. Furthermore, the next-nearest-neighbor hopping $t_c$ shifts the value of $k^{\ast}$ away from zero (cf. Fig \ref{fig8}(a)).
 
We also consider parameters with $t_a=t_b=-1$ and $\gamma=4$, and investigate the slowest decay mode $k^*$ in the purely imaginary spectrum regime  (cf. Fig. \ref{fig8}(b)). The $t^{*}$ consists of 
\begin{equation}
  \begin{cases}
    t_1^{\ast}=\frac{1}{2\sqrt{12 - t_c^2-t_d^2+4 t_c + 4 t_d - 2 t_c t_d}}, & \text{if $k^{\ast}=0$}.\\
    t_2^{\ast} & \text{if $k^{\ast}\neq 0$}.
  \end{cases}
\end{equation}
where $t_2^{\ast}$ is defined as: 
\begin{widetext}
    \begin{equation}
        t_2^{\ast}=\begin{cases}
            \frac{1}{2\sqrt{14-t_c^2-t_d^2 - 2t_c - 2t_d +  \alpha\beta/(3 t_c t_d) + (2\alpha^3+3\alpha^2 \sqrt{3 t_c t_d\beta+(t_c+t_d)^2})/(27 t_c^2 t_d^2)}}, & \text{if $k^{\ast}=\arccos{(\frac{t_c+t_d-\sqrt{9t_c^2 t_d^2+3t_c^2 t_d + 3 t_c t_d^2+t_c^2+t_d^2-t_c t_d}}{6t_c t_d})}$}.\\
            \frac{1}{2\sqrt{14-(t_c+t_d)^2-2(t_c+t_d)-(t_c+t_d-1)^2/(4(t_c+t_d))}}& \text{if $k^{\ast}=\arccos{(-\frac{t_c+t_d-1}{4(t_c +t_d)})}$}.
        \end{cases}
    \end{equation}
\end{widetext}
Here, we define $\alpha = t_c + t_d - \sqrt{3 t_c t_d\beta+(t_c+t_d)^2}$, $\beta = 3 t_c t_d + t_c + t_d - 1$.

We find similar results that the first and second derivatives of $t^{\ast}$ with respect to $t_c$ exhibit a transition signal when the saturation time switches from $t_1^{\ast}$ to $t_2^{\ast}$ (blue lines in Figs. \ref{fig8}(b), (d), and (e)). Furthermore, a crossover occurs within the phase regime corresponding to $t_2^{\ast}$ (green lines in Figs. \ref{fig8}(b), (d), and (e)). The first and second derivatives with respective to $t_c$ show signals near $t_c=0$ or $t_d=0$ but remain continuous throughout the range. This behavior stands in contrast to the transition observed from $t_1^{\ast}$ to $t_2^{\ast}$.

\subsection{Saturation of entanglement entropy}\label{sec5B}
Similar saturation behavior can also be observed in the long-time entanglement entropy. In this subsection, we numerically investigate the relation between long-time half-chain entanglement entropy and phase transitions, as well as the saturation dynamics of the entanglement entropy within the dynamical phases revealed by the spread complexity above.

We consider a partition of the global system into two subsystems $A$ and $B$ with equal length $L$. The von Neumann entropy of subsystem $A$ is given by
\begin{equation}
    S(t) = -\Tr[\rho_{A}(t)\ln{\rho_{A}(t)}],
\end{equation}
where $\rho_{A}(t) = \Tr_{B}[\ket{\psi(t)}\bra{\psi(t)}]$ is the reduced density matrix of subsystem $A$.  We estimate the asymptotic long-time value $\bar{S}$ by averaging the entanglement entropy over a suitable time window, similar to Eq. (\ref{eq:time_average}).

We use the numerical method in Refs. \cite{10.21468/SciPostPhys.7.2.024,PhysRevLett.126.170602,10.21468/SciPostPhysLectNotes.82,PhysRevX.13.021007} to investigate the dynamics of entanglement entropy. We prepare the initial state as the reference state in Eq. (\ref{eq:gs}) in the real space, which is a Gaussian state at half-filling. Since the Hamiltonian (cf. Eq. (\ref{eq:real_ham})) is quadratic, the evolved state remains to be a Gaussian state. Thus, the evolved state can be represented as
\begin{equation}
    \ket{\psi(t)}=\prod_{n=1}^{L}\Big(\sum_{j=1}^{2L}U_{jn}(t)c^\dagger_{j}\Big)\ket{0},
\end{equation}
where $c^\dagger_{j,A}=c^\dagger_{2j}$ and $c^\dagger_{j,B} = c^\dagger_{2j+1}$ and $U_{jn}(0)$ can be obtained by the Fourier transformation of Eq. (\ref{eq:gs}) and satisfy $U^\dagger U=I$.
The time evolution of $U(t)$ for the time interval $\Delta t$ is given by 
\begin{align}
    \begin{split}
        \ket{\psi(t+\Delta t)}&\propto e^{-iH\Delta t}\ket{\psi(t)}\\
        &=\prod_{n=1}^{L}\Big(\sum_{j=1}^{2L}[e^{-i\mathbf{h}\Delta t} U]_{jn}c_j^\dagger\Big)\ket{0},
    \end{split}
\end{align}
where $\mathbf{h}$ is the single-particle Hamiltonian $H=\sum^{2L}_{ij}c_i^\dagger \mathbf{h}_{ij} c_j$ (cf. Eq. (\ref{eq:real_ham})). To restore the normalization condition $\braket{\psi(t)}{\psi(t)}=1$, we perform the QR decomposition $e^{-i\mathbf{h}\Delta t} U=QR$, where $Q$ is an $2L\times L$ matrix satisfying $Q^\dagger Q=1$ and $R$ is an upper triangular matrix. The $2L\times L$ matrix $U(t+\Delta t)=Q$. In our numerical calculation, $\Delta t = 0.1$, $L=100$ and the step size of the parameters is set to $0.04$ to reduce the calculation cost. 

The matrix $U(t)$ contains all information about the quantum dynamics. In particular, the $2L\times 2L$ correlation matrix $C_{ij}(t):=\bra{\psi(t)}c_i^\dagger c_j\ket{\psi(t)}$ is obtained as $C(t)=[U(t)U^\dagger(t)]^T$. To calculate the half-chain entanglement entropy $S$, we can diagonalize the $L\times L$ submatrix $C$ and obtain its eigenvalues $\lambda_n$'s. Then the von Neumann entanglement entropy is given as \cite{Calabrese_2005,PhysRevA.78.010306,Alba_2009} 
\begin{equation}
    S=-\sum_{i=1}^{L}[\lambda_i \log \lambda_i + (1-\lambda_i)\log(1-\lambda_i)].
\end{equation}

The long-time entanglement entropy $\bar{S}$ exhibits similar behavior as the long-time spread complexity $\bar{C}$ (cf. Fig \ref{fig4}(c)-(d)). The derivative of the long-time entanglement entropy detects the phase transitions through prominent jumps/ drops in both the Hermitian and non-Hermitian cases.

For the relaxation behavior of entanglement entropy, although the saturation value of entanglement entropy is dependent on the initial state and parameters in the Hamiltonian (cf. the inset of Fig. \ref{fig7}(b)), the saturation time exhibits a similar non-monotonic behavior, suggesting that the entanglement entropy captures the same physics as the spread complexity (cf. Fig. \ref{fig7}(b)). We use $|\Delta S|=|S(t)-S(t\rightarrow\infty)|$ as an indicator of saturation and characterize how $|\Delta S|$ approaches zero. In practice, $S(t \rightarrow \infty)$ is obtained by taking the time average over a sufficiently long evolution period to ensure convergence, and we use $\epsilon=10^{-10}$ as the saturation marker to pinpoint the saturation times $\tilde{t}_{\epsilon=10^{-10}}^{EE}$ (cf. Fig. \ref{fig7}(b)). The numerically obtained entanglement entropy saturation time $\tilde{t}_{\epsilon}^{EE}$ is found to follow the same relation as Eq. (\ref{eq:t_epsilon}) (cf. Fig. \ref{fig7}(d)). By fitting the numerically obtained saturation time, we establish this relationship holds upon substituting the prefactor $\mathcal{C}_1$ with a fitted coefficient $\mathcal{C}_2$. This relation is insensitive to the choice of $\epsilon$ (cf. Fig. \ref{fig7}(e)). Furthermore, by extrapolating the saturation time $\tilde{t}_{\epsilon}^{EE}$ to the $\epsilon\rightarrow 0$ limit, the numerically obtained $\epsilon$-independent saturation time $\tilde{t}^{\ast}_{EE}$ also approaches the analytical result $t^{\ast}$ obtained from Eq. (\ref{eq:t_epsilon_indep}) (cf. Fig. \ref{fig7}(f)). Therefore, the similar behavior exhibited by the entanglement entropy and spread complexity confirms that the dynamical phases are also manifest in entanglement entropy, thereby providing a more accessible quantity for experimental observation.

\section{conclusions }\label{sec6}
We study the phase diagram regarding the spectral properties of the non-Hermitian extended SSH model with next-nearest-neighbor hoppings and an imaginary staggered chemical potential. In the Hermitian case under equilibrium, the number of gap-closing momenta at the topological phase transitions is equal to the absolute difference of the winding number of the phases on the two sides of the phase boundary. In the presence of small non-Hermiticity, we find that the exceptional points emerge in pairs from the gap-closing momenta in the Hermitian limit.

We then use Krylov spread complexity and entanglement entropy to study these phase transitions under two dynamical protocols: (i) preparing the ground state of the non-Hermitian model through unitary evolution from an initial reference state, and (ii) the long-time evolution under the non-Hermitian Hamiltonian from the initial reference state. For the first protocol, the spread complexity is a robust detector, where its first derivative shows discontinuities at the phase transition points.  For the second protocol, the first derivatives of both long-time spread complexity and entanglement entropy are also able to detect the phase transitions. However, minor peaks or valleys are also observed within a phase, which suggests Krylov complexity in the first protocol may be a better indicator of transitions.

We further introduce the mode-resolved complexity to unravel the detailed structure of the transitions. The mode-resolved complexity identifies the critical momenta that correspond to the transitions, and how these critical momenta evolve as the driving parameter varies. We quantify the similarity in the mode-resolved complexity's distributions by fidelities. The fidelity with respect to the driving parameter exhibits sharp changes at the transitions and accurately locates the phase boundaries in both Hermitian and non-Hermitian models. On the other hand, the fidelity with respect to the momenta detects the gap-closing momenta and their evolution with the model parameter in the Hermitian limit, while in the non-Hermitian case it identifies the characteristic momenta marking the onset and offset of gap-closings.

In the saturation regime where the system possesses a purely imaginary spectrum, we find that the dynamical phases existing in the non-Hermitian SSH model \cite{6lvg-7qdn} persist with the additional hoppings and can be further tuned by these additional parameters. The dynamical phases are distinguishable from the saturation time of the spread complexity and the entanglement entropy, in which both are related to the inverse of the energy gap.

Future studies could expand on the above findings in several directions. First, for more general models with a complex spectrum (where eigenvalues possess both real and imaginary parts) \cite{PhysRevB.111.174207}, it has been shown that non-trivial dynamical features are governed not only by the slowest decay mode but also by a broader set of modes, including a continuum of them. In such regimes, how does the mode-resolved complexity behaves, and whether the entanglement entropy can effectively capture these phase transitions requires further investigation. Additionally, identifying order parameters to characterize the equilibrium and the dynamical phases in the non-Hermitian model remains an open and interesting question. Furthermore, regarding the physics of biorthogonal dynamical quantities, there is room for deeper exploration. Beyond utilizing explicit normalization for time-dependent probabilities and observables, one could adopt the framework of biorthogonal quantum mechanics by defining an associated state to reformulate the transition probability \cite{Brody:2013axr}. Alternatively, one could employ the quantum metric framework wherein the Hilbert space is non-stationary and endowed with a non-trivial time-dependent metric \cite{PhysRevA.93.042114,e22040471}. Finally, investigating other measures of quantum complexity \cite{RevModPhys.91.025001} represents another promising direction.

\section*{acknowledgments}
We acknowledge financial support from Research Grants Council  of  Hong  Kong  (Grant  No. CityU 11318722), and City University of Hong Kong (Grants No. 9610438, 7020156).

\bibliography{apssamp}

\end{document}